\documentclass[twocolumn,prl,showpacs]{revtex4}
\usepackage{amsmath}
\usepackage{amssymb}
\usepackage{graphicx}
\usepackage{dcolumn}
\usepackage{bm}
\usepackage{float}
\usepackage{subfigure}
\usepackage{epsfig}
\usepackage[colorlinks, linkcolor=blue, anchorcolor=blue,citecolor=blue]{hyperref}

\begin{document}

\title{Nonlinear effects in the laser-assisted scattering of a positron by a muon }

\author{Wen-Yuan Du}\email{pmcnpx@mail.ustc.edu.cn}

\affiliation{Department of Modern Physics, University of Science and Technology of China, Hefei, 230026, P.R. China}

\author{Bing-Hong Wang}\email{bhwang@ustc.edu.cn}

\affiliation{Department of Modern Physics, University of Science and Technology of China, Hefei, 230026, P.R. China}

\author{Shu-Min Li}

\affiliation{Department of Modern Physics, University of Science and Technology of China, Hefei, 230026, P.R. China}

\begin{abstract}
The scattering of a positron by a muon in the presence of a linearly polarized laser field is investigated in the first Born approximation. The theoretical results reveals: 1) at large scattering angle, an amount of multiphoton processes take place in the course of scattering. The photon emission processes predominant the photon absorption ones. 2) Some nonlinear phenomena about oscillations, dark angular windows, and asymmetry can be observed in angular distributions. We analyze the reason giving rise to ¡°dark windows¡± and geometric asymmetry initially noted in the potential scattering. 3) we also analyze the total differential cross section, the result shows that the larger the incident energy is, the smaller total differential cross section is. The reason of these new results are analysed.
\end{abstract}


%32.80.-t Photon interactions with atoms (see also 42.50.-p Quantum optics)
%32.80.Fb Photoionization of atoms and ions
%33.80.Wz	Other multiphoton processes
%34.50.Rk Laser-modified scattering and reactions
%34.80.Qb Laser-modified scattering
\maketitle

\section{Introduction}

Nonlinear effects in the processes of interaction of an electron with a nucleus, or with other charged particles in the some optical field have become the focus of the study. The earlier study gives the two parameters which governed these nonlinear effects~\cite{Ni79}. The first one is the classical relativistic-invariant parameter of intensity of the wave $\eta_{e}$, which denotes $\eta_{e}=\frac{eF_{0}}{m_{e}c\omega}$, here $e$ and $m_{e}$ are the charge and the mass of an electron; $c$ is the light velocity; $F_{0}$ and $\omega$ are the strength and the frequency of the electric field in the wave. The second one is the quantum Bunkin-Fedorov parameter $\gamma_{0e}=\eta_{e}\frac{m_{e}\upsilon_{e}c}{\hbar\omega}$, which determines the multiplicity of a multiphoton process~\cite{B65,D67,Fed91}. Note that a quantum Bunkin-Fedorov parameter is the principal parameter of the multiphoton processes in the nonresonant scattering(see~\cite{Ned07,Pad09}), In the case of $\gamma_{0e}\ll1$, single photon exchange channel provides the main contribution to the scattering cross-section in the external field presence. When $\gamma_{0e}\gtrsim1$, a process of nonresonant scattering becomes a nonlinear (multiphoton) one with respect to the external field. In the work~\cite{Pad09}, the authors found the cross-section of nonresonant scattering of relativistic electron by relativistic muon in the moderately strong quasimonochromatic laser field differs substantially from the corresponding cross-section of the process in the field of plane monochromatic wave.
Therefore,  the multiphoton processes in the relativistic regime attracted more attention. The multiphoton processes of potential scattering in the relativistic regime has been investigated in~\cite{MK}. Nonlinear Compton scattering induced by a linearly polarized laser field with high-velocity electrons has been studied by~\cite{PP12}. Moreover, many authors have been enthusiastically investigating some other interesting multiphoton processes, such as laser-assisted bremsstrahlung~\cite{EL,SS}, and pair production~\cite{CK}.

In this paper we investigate the relativistic scattering of a positron by a muon in the presence of a linearly polarized laser field of a medium intensity in frame of the Born approximation. After numerical calculation and mathematical analysis, the nonlinear effects in the scattering process are revealed, such as multiphoton processes, cut off characteristic phenomena, as well as dark angular windows and asymmetry effects~\cite{Kam99,PP02}. We also analyse the difference regarding to the multiphoton processes between the positron-muon scattering and the electron-muon scattering. In contrast to the electron-muon scattering, we find that the photon emission processes predominant the photon absorption ones for positron-muon scattering. For the total differential cross section, the result shows that the larger the incident energy is, the smaller total differential cross section is.

The paper is arranged as follows: in Sec. II, we derive the expression of the differential cross section of a positron by a muon in a linearly polarized laser field. In Sec. III, numerical results are presented and the dependence of the differential cross section on scattering angles, azimuthal angles and various types of parameters are discussed. Section IV is devoted to a brief summary and conclusions. Natural units $\hbar=c=1$ and Minkowski metric tensor $g^{\mu\nu}=diag(1,-1,-1,-1)$ are used throughout this paper.

\section{theory of the scattering $e^++\mu^-\rightarrow e^++\mu^-$}

 We begin with the process of laser-assisted scattering
 of a positron by a muon in the lowest order of perturbation theory,
 it can be described by the lowest Feynman
diagrams. The Scattering amplitude can be written as
\begin{eqnarray}
S_{fi}=i e^2 \int d^4 x \int d^4 y \bar{{\psi}}_{e^{+}}^{i}(x)
\gamma_{\mu} {\psi}_{e^{+}}^{ f}(y) \nonumber\\
\times D^{\mu \nu}(x-y) \bar{{\psi}}_{\mu ^{-} }^{f} (x) \gamma_{\nu}
{\psi}_{\mu ^{-} }^{i}(y),
\end{eqnarray}
here, $D_F(x-y)=\int {\frac {d^4 k'}{(2 \pi)^4}}{\frac {-4 \pi
}{k'^2+ i \varepsilon }} e^{-i k'(x-y)}$ is the Feynman propagator
for photons~\cite{Greiner}.
The wave function of the muons can be described
by Dirac function when normalized in volume $V$,
\begin{eqnarray}
{\psi}_{\mu ^{-} }^{i} (x)={\frac {1}{\sqrt{2 E_{i} V}}}
u(P_{i},S_{i}) e^{-i P_{i}x}.
\end{eqnarray}
\begin{eqnarray}
{\psi}_{\mu ^{-} }^{f} (y)={\frac {1}{\sqrt{2 E_{f} V}}}
u(P_{f},S_{f}) e^{-i P_{f}y}.
\end{eqnarray}
The laser assisted electron wave function could be described
by Dirac-Volkov state~\cite{Volkov}
\begin{eqnarray}
{\psi}_{e^{+}}^{i}(x) ={\frac {1}{\sqrt {2 Q_{i} V}}} (1 - {\frac {e
k\!\!\!/ A\!\!\!/}{2k p_i}}) {\nu}(p_i,s_i) {\rm
e}^{i{S_{p_{i}}[\phi(x)]}},
\end{eqnarray}
\begin{eqnarray}
{\psi}_{e^{+}}^{ f}(y) ={\frac {1}{\sqrt {2 Q_{f} V}}} (1 -{\frac {e
k\!\!\!/ A\!\!\!/}{2k p_f}}) {\nu}(p_f,s_f) {\rm
e}^{i{S_{p_f}[\phi(y)]}},
\end{eqnarray}
with
\begin{eqnarray}
{S}_{p_{i,f}}[\phi(x)]=p_{i,f}x \nonumber\\
+ \int_0^\phi \left ({e \frac
{p_{i,f} A(\phi ')}{k p_{i,f}}}-{e^2 \frac {A^2 (\phi ')}{2 k
p_{i,f}}} \right ) d \phi ',
\end{eqnarray}
where, the under symbol $i$ and $f$ mark the incoming and outgoing
positron (muons) wave function.The $\nu_{p_{i,f},s_{i,f}}$ are the
spinors of positrons satisfying $\sum\limits_{s} \nu (p,s) \bar {\nu} (p,s)=p \!\!\!/-m_{e^{+}}$.

For mathematical simplicity, the laser field is taken to be a
monochromatic plane wave of linearly polarization with the classical
four-potential.
\begin{eqnarray}
{\bm A}(x)={\bm a} {\epsilon}^{\mu} {\cos\phi} ,\label{eqA}
\end{eqnarray}
where ${\bm a}={\boldsymbol{\mathcal E}}_0/\omega$,
${\boldsymbol{\mathcal E}}_0$ is the strength of a laser
field.${\phi}=kx$, where $k=(\omega,{\bm k})$, $\omega$ and ${\bm k}$
are the frequency and the wave number, respectively. $\epsilon$ is the
polarization four-vector satisfying $\epsilon k=0$ and
${\epsilon}^2=-1$.

The laser assisted kinetic momentum
of the positron is called effective momentum $q^{\mu}$, with the
form
\begin{eqnarray}
q^{\mu}=p^{\mu}-{\frac {e^2 \overline A^2}{2 (k p)}} k^{\mu}.
\end{eqnarray}
Its square is $q^2=m_*^{2}=m^2_{e^{+}}+e^2 a^2/2$, where $m_*$ acts as
an "effective mass" of the positron in the laser field.

The space-time integrations in Eq.(1) can be performed by the
standard method of Fourier series expansion, using the generating
function of the generalized Bessel functions. Completing the coordinate integral in Eq. (1) by
the ansatz of Bessel expansion
$$ \left\{
\begin{aligned}
1 \\
\cos\phi \\
\cos2\phi \\
\end{aligned}
\right\} \times e^{i\xi\sin\phi-i\eta\sin2\phi}
$$
$$
=\sum\limits_{n} e^{in\phi} \times \left\{\begin{aligned}
B_n(\xi,\eta)\\
\frac{1}{2}[B_{n+1}(\xi,\eta)+B_{n-1}(\xi,\eta)]\\
\frac{1}{2}[B_{n+2}(\xi,\eta)+B_{n-2}(\xi,\eta)]\\
\end{aligned}
\right\}
$$
where $B_n(\xi,\eta)$ stands for the generalized Bessel function,

\begin{eqnarray}
B_n(\xi,\eta)=\sum\limits_{\lambda=-\infty}^{\infty}J_{n-2\lambda}(\xi)J_{\lambda}(\eta),
\end{eqnarray}

We obtain
\begin{eqnarray}
S_{fi} = -i {\frac {(2 \pi )^44 \pi e^2}{4 V^2 \sqrt { Q_i Q_f E_i
E_f}} }\sum\limits_{l=-\infty}^{\infty} {\frac { M_l
}{(P_i-P_f)^2 +i \epsilon }} \nonumber\\
\times \delta^4 (q_f-q_i +P_f-P_i+l k),
\end{eqnarray}
In which,
\begin{eqnarray}
M_l=\bar{\nu}(p_i,s_i) \Gamma^{\mu} \nu(p_f,s_f)\bar{u}(P_f,S_f)
\gamma_{\mu} u(P_i,S_i) ,
\end{eqnarray}
with
\begin{eqnarray}
\Gamma^{\mu}=\Delta_0 \gamma^{\mu}+\Delta_1  k \!\!\!/
\epsilon\!\!\!/ \gamma^{\mu} +\Delta_2 \gamma^{\mu}k\!\!\!/\epsilon\!\!\!/
+\Delta_3 k\!\!\!/\epsilon \!\!\!/
\gamma^{\mu}k \!\!\!/ \epsilon\!\!\!/,
\end{eqnarray}
where
\begin{eqnarray}
\Delta_0&=&B_l(\xi,\eta),\\
\Delta_1&=&\frac{ea[B_{l-1}(\xi,\eta)+B_{l+1}(\xi,\eta)]}{4kp_i },\\
\Delta_2&=-&\frac{ea[B_{l-1}(\xi,\eta)+B_{l+1}(\xi,\eta)]}{4kp_f},\\
\Delta_3&=-&\frac{e^2 a^2[2 B_l(\xi,\eta)
+B_{l-2}(\xi,\eta)+B_{l+2}(\xi,\eta)]}{16 kp_i kp_f},
\end{eqnarray}
with
\begin{eqnarray}
\xi = ea \left ({\frac{\epsilon p_f}{k p_f}}-{\frac {\epsilon p_i}{k
p_i}}\right),
\end{eqnarray}
\begin{eqnarray}
\eta = e^2 a^2\left ({\frac{1}{8 k p_f}}-{\frac {1}{8 k
p_i}}\right).
\end{eqnarray}

Here, we do not consider the polarization effects of the positron and muon. For calculating the total scattering cross section , we sum over the
final spin states and averaged over the initial spin states. On
account of that the muon have a spin $1/2$ such as the electron. The
total cross section could be obtained as
\begin{eqnarray}
\bar \sigma= V \int \frac {d^3 {\bm p}_f}{(2 \pi)^3} V \int \frac
{d^3 {\bm P}_f}{(2 \pi)^3} \frac{1}{4} \sum\limits_{s_i s_f S_i S_f}
\frac{|S_{fi}|^2 }{V T J_{\nu}V^{-1}}\nonumber\\
=\sum\limits_l\frac{e^2}{4 \sqrt{(P_iq_i)^2-m^2_{\mu} m^2_{e^{+}}}}
\int\frac{d^3 {\bm q}_f}{2Q_f} \int\frac{d^3 {\bm P}_f}{2 E_f}\nonumber\\
\times\delta^4(q_f-q_i+P_f-P_i+l k) \
\frac{1}{4}\sum\limits_{s_i,s_f,S_i,S_f}
\frac{{|M_l|}^2}{(P_i-P_f)^4}.
\end{eqnarray}
Here $J_{\nu}$ is the incoming current in the laboratory system
given by $\sqrt{(P_i q_i )^2-m^2_{\mu} m^2_{e^{+}} } /(E_1 Q_1V)$. The
total differential cross section can be decomposed into a series of
discrete partial differential cross sections for different numbers
of photon transfer,
\begin{eqnarray}
\frac{d\bar\sigma_{tot}}{d\Omega}=\sum\limits_{l=-\infty}^{\infty}\frac{d\bar\sigma_l}{d\Omega}.
\end{eqnarray}

We want to calculate the total differential cross section for
positron scattering into a given solid-angle element $d\Omega$
centered around the scattering angle $\theta$. Therefore the
differential quantity has to be integrated over all momentum
variables expect for the direction of $dp_f$. We use the Jacobin determinant $J$ to get
the involve transformation of integration variable between the
free positron 4-momentum $p^{\mu}$ and the positron 4-momentum
shifted by the presence of the external field $q^{\mu}$,
\begin{eqnarray}
d^{4}p=J d^{4}q ;
J=\left|
\begin{array}{cccc}
\frac{\partial p^{0}}{\partial q^{0}} \frac{\partial p^{0}}{\partial q^{1}} \frac{\partial p^{0}}{\partial q^{2}} \frac{\partial p^{0}}{\partial q^{3}}\\
\frac{\partial p^{1}}{\partial q^{0}} \frac{\partial p^{1}}{\partial q^{1}} \frac{\partial p^{1}}{\partial q^{2}} \frac{\partial p^{1}}{\partial q^{3}}\\
\frac{\partial p^{2}}{\partial q^{0}} \frac{\partial p^{2}}{\partial q^{1}} \frac{\partial p^{2}}{\partial q^{2}} \frac{\partial p^{2}}{\partial q^{3}}\\
\frac{\partial p^{3}}{\partial q^{0}} \frac{\partial p^{3}}{\partial q^{1}} \frac{\partial p^{3}}{\partial q^{2}} \frac{\partial p^{3}}{\partial q^{3}}\\
\end{array}\right|.
\end{eqnarray}
In finding the Jacobian of this transformation we must make use of a coordinate system in which
the external field photons propagate along the y-axes, $k^{\mu} = (\omega, 0, \omega, 0)$ , with the result that the free
positron momentum can be written
\begin{eqnarray}
p^{\mu}=q^{\mu}+{\frac {e^2 a^2}{2\omega(q^{0}-q^{2})}} k^{\mu}.
\end{eqnarray}
The Jacobian J then reduces to
\begin{eqnarray}
J=\left|
\begin{array}{ccccccc}
\frac{\partial p^{0}}{\partial q^{0}}& 0& \frac{\partial p^{0}}{\partial q^{2}}& 0\\
0 &                                 1& 0    &                        0\\
\frac{\partial p^{2}}{\partial q^{0}}& 0& \frac{\partial p^{2}}{\partial q^{2}}& 0\\
0 &                              0& 0    &                              1\\
\end{array}\right|=\frac{\partial p^{0}}{\partial q^{0}}\frac{\partial p^{2}}{\partial q^{2}}- \frac{\partial p^{0}}{\partial q^{2}}\frac{\partial p^{2}}{\partial q^{0}}=1.
\end{eqnarray}
Therefore, the volume element can be written as $d^3 {\bm q}_f=|{\bm q}_f|^2 d|{\bm q}_f| d\Omega$, and
$|{\bm q}_f|d|{\bm q}_f|=Q_2 dQ_2$, and with the help of $\frac{d^3
P}{2E}=\int \limits_{l=-\infty}^{\infty} d^4 P
\delta(P^2-M^2_0)\Theta(P_0)$ and $ \int dx f(x)
\delta[g(x)]=[f(x)/|g'(x)|]|_{g(x)=0}$. Thus by integrating over $d
|{\bm q}_f|$ and $d^4 P_f$, the partial differential cross section
can be written as
\begin{eqnarray}
& &\frac{d\sigma_l}{d\Omega}= \frac{e^4}{8 m_{\mu}}\frac{|{\bm
q_f}|^2}{|{\bm q_i}|} \frac{1}{(P_i-P_f)^4 Q_f \,|g'(|{\bm q_f}|)|}
\sum\limits_{s_i s_f S_i
S_f}|M_l|^2\nonumber\\
& &|_{P_f=P_i+q_i-q_f-lk;|{\bm
q_f}|=\sqrt{\frac{A^2C^2}{(A^2-B^2)^2}-\frac{C^2-B^2m^2_*}{A^2-B^2}}
-\frac{AC}{A^2-B^2}},
\end{eqnarray}
with
\begin{eqnarray}
g'(|{\bm q_f}|)=-2m_{\mu}|{\bm q_f}|/Q_f-2Q_i |{\bm q_f}|/Q_f\nonumber\\
+2|{\bm q_f}|/Q_f l \omega  +2|{\bm q_i}| \cos{\theta}
-2l\omega\sin{\theta} \sin{\varphi},
\end{eqnarray}
and $A=|{\bm q_i}|\cos{\theta}-l \omega
\sin{\theta}\sin{\varphi}$,$B=m_{\mu}+Q_1-l\omega$,$C=m^2_*+m_{\mu}
Q_i-m_{\mu}l\omega-Q_il\omega+l{\bm q_i}\cdot{\bm k}$, and $\theta$
is the scattering angle of the incoming positron, and $\varphi$ is the
polarization angle of the outgoing positron, and
\begin{eqnarray}
& &\sum\limits_{s_{i,f},S_{i,f}}|M_l|^2=\frac{1}{16m^2_{\mu} m^2_{e^{+}}}
Tr[({p\!\!\!/}_f-m_{e^{+}})\Gamma^{\mu}({p\!\!\!/}_i-m_{e^{+}})\Gamma^{\nu}]\nonumber\\
& &Tr[({P\!\!\!/}_f+m_{\mu})\gamma_{\mu}({P\!\!\!/}_i+m_{\mu})\gamma_{\nu}],
\end{eqnarray}
The trace work can be performed with the help of FEYNCALC.

\section{NUMERICAL RESULTS AND DISCUSSION}
In this section,we present and discuss the numerical result for cross section of the scattering of a muon and a positron in the linearly polarized laser field. We set the incident electron momentum $p_i$ along the $z$ axis and the incident positron momentum $-p_{i}$. For simplifying the calculation process, we use a fixed target in the laboratory frame. We evaluate the cross section in the rest frame of the incoming muon with the initial energy  $E_{i}=m_{\mu}$. Considering the relativistic effect, we set the incident positron kinetic energy as $10^{6}eV$.

The energy conservation relation derived from the $\delta$ function is $Q_{2}(m_{\mu}+Q_{i})-(q_{f}\cdot q_{i}+lq_{f}\cdot k)=m^2_{*}+m_{\mu}Q_{i}$. In the limit case of $Q_{i}\ll m_{\mu}$, The energy conservation relation here becomes $Q_{f}=Q_{i}+lk$. $lk$ is the photon energy transmission from laser fields ($l>0$ for emission, and $l<0$ for absorption). That is the multiphoton effect for elastic scattering: redistribution of the total energy and momentum between two participating particles and external field.

\begin{figure}[htp]

\centering
\subfigure[]{
\label{fig:subfiga}
\includegraphics[bb=56 73 722 577,width=2.6in]{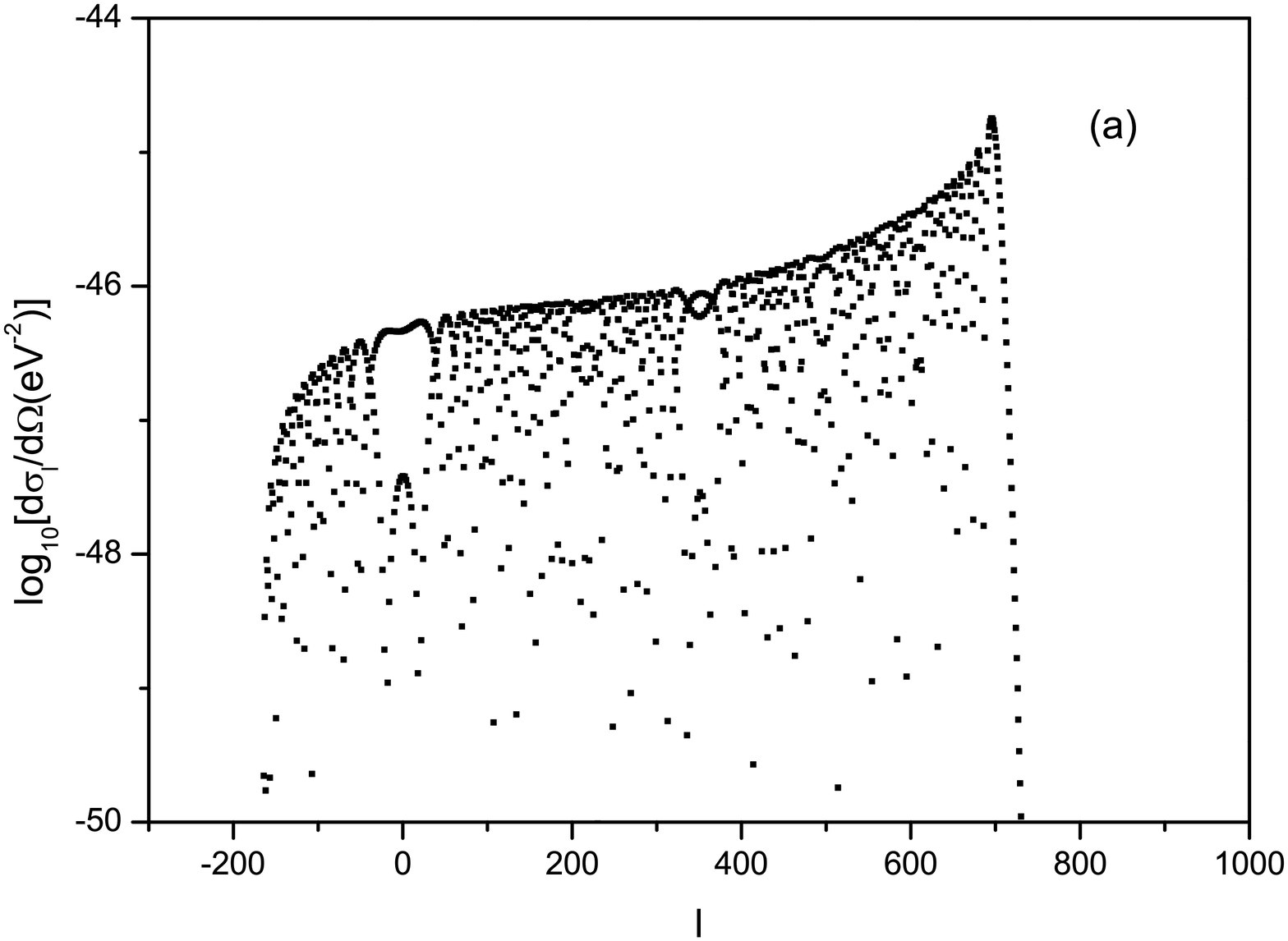}}

\vspace*{0.1in}
\subfigure[]{
\label{fig:subfigb}
\includegraphics[bb=56 73 722 577,width=2.6in]{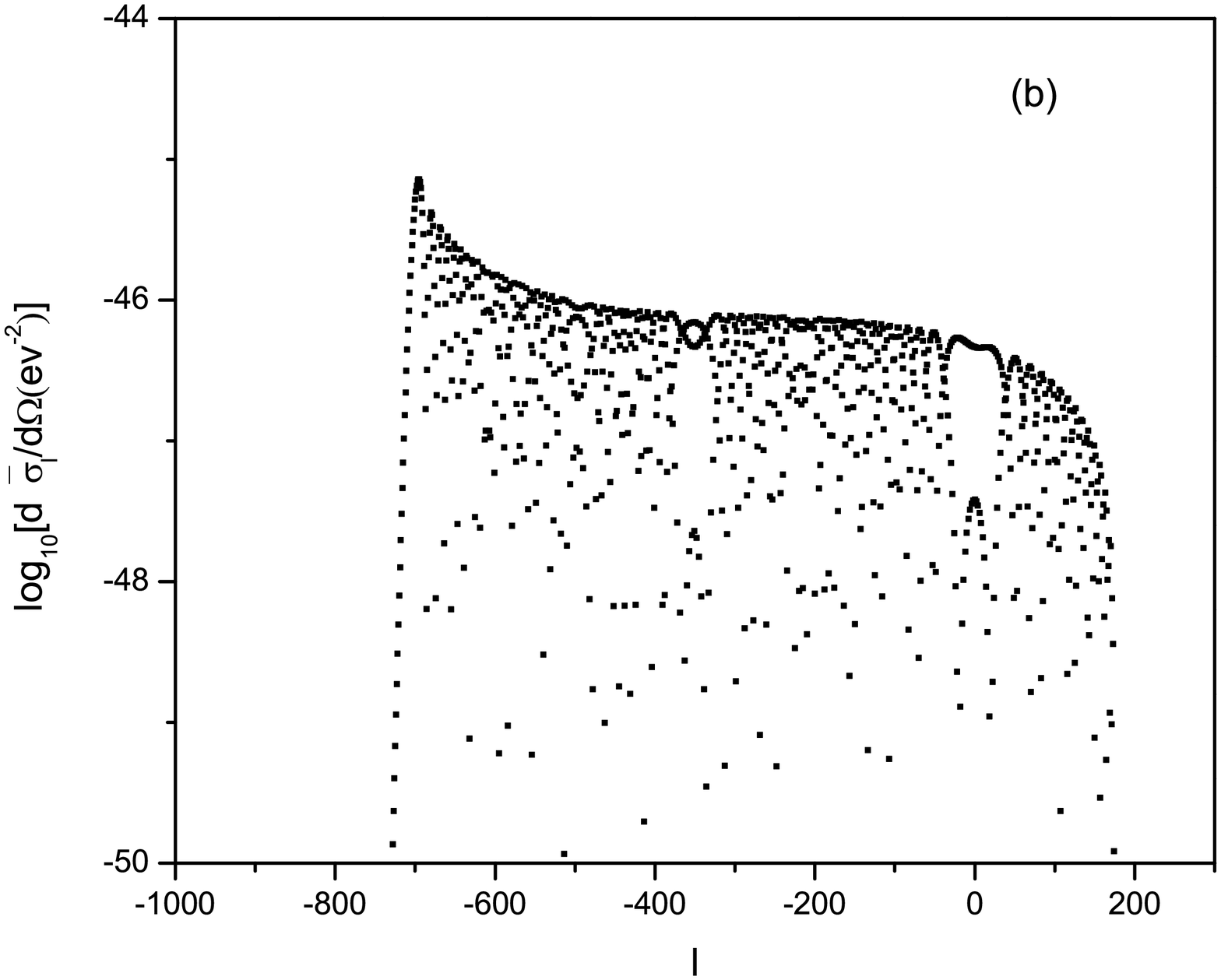}}

\caption {\label{CgL} The multiphoton cross section for the laser-assisted scattering of a positron (electron) by a muon at an impact energy of $T_{i}=0.511MeV$ as a function of the number of the involved photons $l$. $(a)$ for the positron, $(b)$ for the electron. The scattering angle: $ \theta=90^{o}$, and
azimuthal angle: $\varphi=0^{o}$ . The laser field is linearly polarized along the incident direction of the positron. The field strength is $\varepsilon_{0}=5.18\times10^{7} V/cm$, and the photon energy $\hbar\omega=1.17eV$.}
\end{figure}
Figure~\ref{CgL} shows the partial differential cross sections $d\overline{\sigma}_{l}/d\Omega$ versus the net photon number $l$ transferred between the colliding system and the laser field. We display the partial differential cross sections at a large scattering angle ($\theta=90^{o}$) both of the positron in Fig.~\ref{fig:subfiga} and the electron in Fig.~\ref{fig:subfigb}. The strength and frequency of the laser field are $\varepsilon_{0}=5.18\times10^{7}V/cm$ and $\omega=1.17eV$. The direction of the field wave vector $k$ is along the $y$ axis, and the laser polarization is chosen to be parallel to the positron (electron) incident momentum along the $z$ axis. We find a large number of photons are exchanged between the laser field and the colliding system, in which multiphoton processes take place. This nonlinear phenomenon perhaps arise from the resonant state of the positron(electron) and the muon formed in the collision process. The result shown in Fig.~\ref{fig:subfiga} that the photon emission processes predominant the photon absorption ones. In this collision process, initial positron and muon formed a hydrogen-like intermediate transient state due to the Coulomb interaction. This transient state is not stable, positron and muon dissociated from this state, therefore emission lots of photons than absorbing. Fig.~\ref{fig:subfigb} shows the opposite result, in which the photon absorption processes predominant the photon emission ones. For the electron, which has the different sign of the charge, due to the Coulomb repulsion, the number of the absorbing photons is larger than the number of the emission photons.
The magnitudes for $d\overline{\sigma_{l}}/d\Omega$ varies in the range of few orders for different $l$. These oscillations take on owing to the periodical
variation behavior of the generalized Bessel function $\textbf{B}(\xi,\eta)$~\cite{Erik}. Furthermore, the contribution of various l-photon processes are cut off at two edges which are asymmetric with respect to $l=0$. The symmetry axis of the generalized bessel function is at $l=-2\eta$. The cutoff for positive $l$ is a consequence of the energy conservation imposed on Eq.(18). On the other hand, the origin of the cutoff for negative values of $l$ can be inferred by the properties of the generalized Bessel function $\textbf{B}(\xi,\eta)$ when its arguments $\xi$ and $\eta$ satisfy the approximate relation $l=\pm|\xi|\pm2|\eta|$. This has already been pointed out in~\cite{Erik}.

Nonlinear phenomena can also be observed in angular distributions: oscillations, dark angular windows, and asymmetry. These can also be better explained from the behavior of the generalized Bessel function. The reason for ¡°dark windows¡± is similar to the cutoff phenomenon that the value of the generalized Bessel function diminishes rapidly to almost zero when its order exceeds its arguments, which causes the partial differential cross section to become so small that such events cannot be collected efficiently in some particular angular regions~\cite{Kam99,FH}.

\begin{figure}[htp]

\centering
\subfigure[]{
\label{fig:subfig:2a}
\includegraphics[bb=37 22 732 592,width=2.6in]{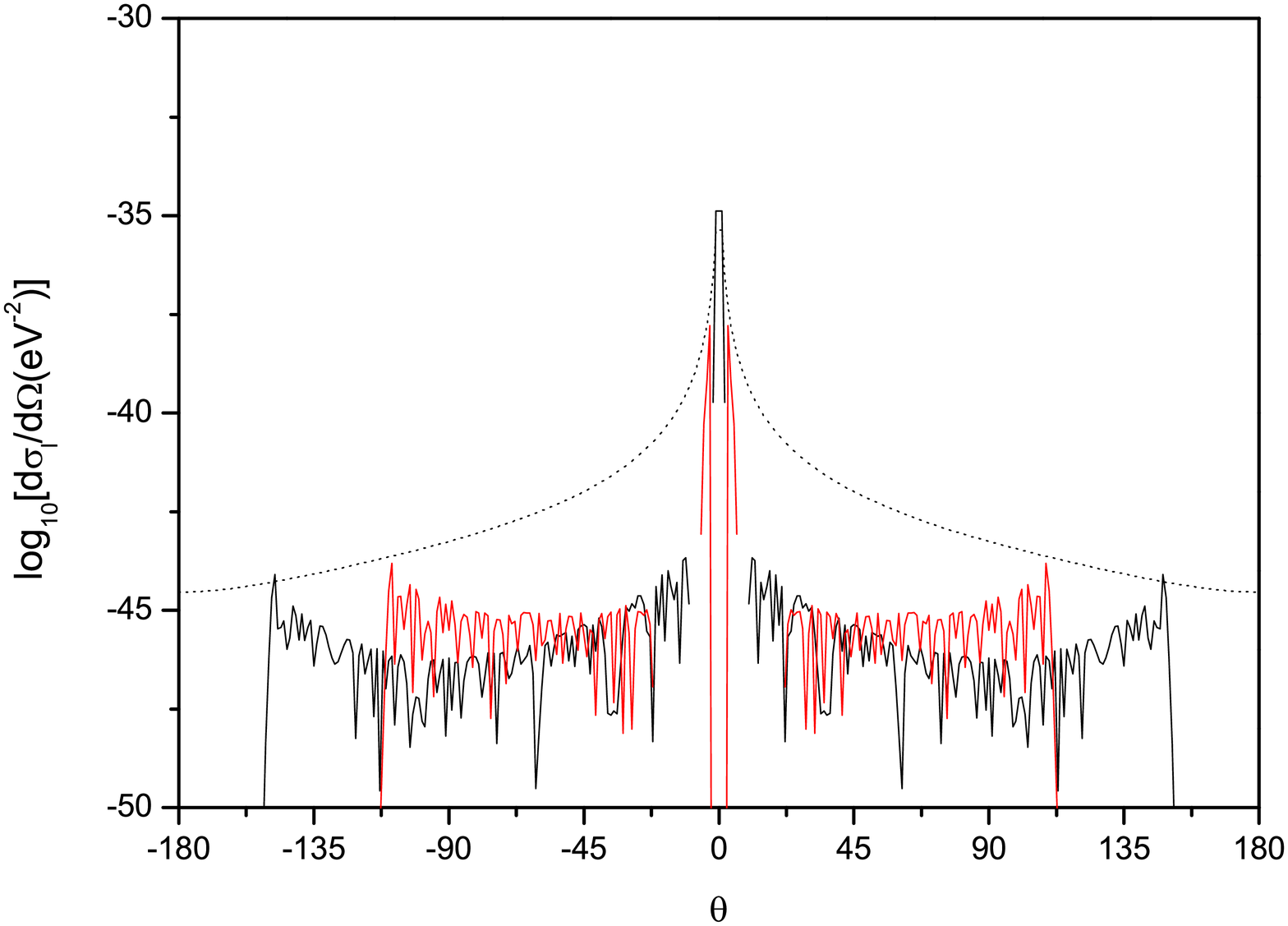}}

\hspace{0.1in}
\centering
\subfigure[]{
\label{fig:subfig:2b}
\includegraphics[bb=56 28 736 594,width=2.6in]{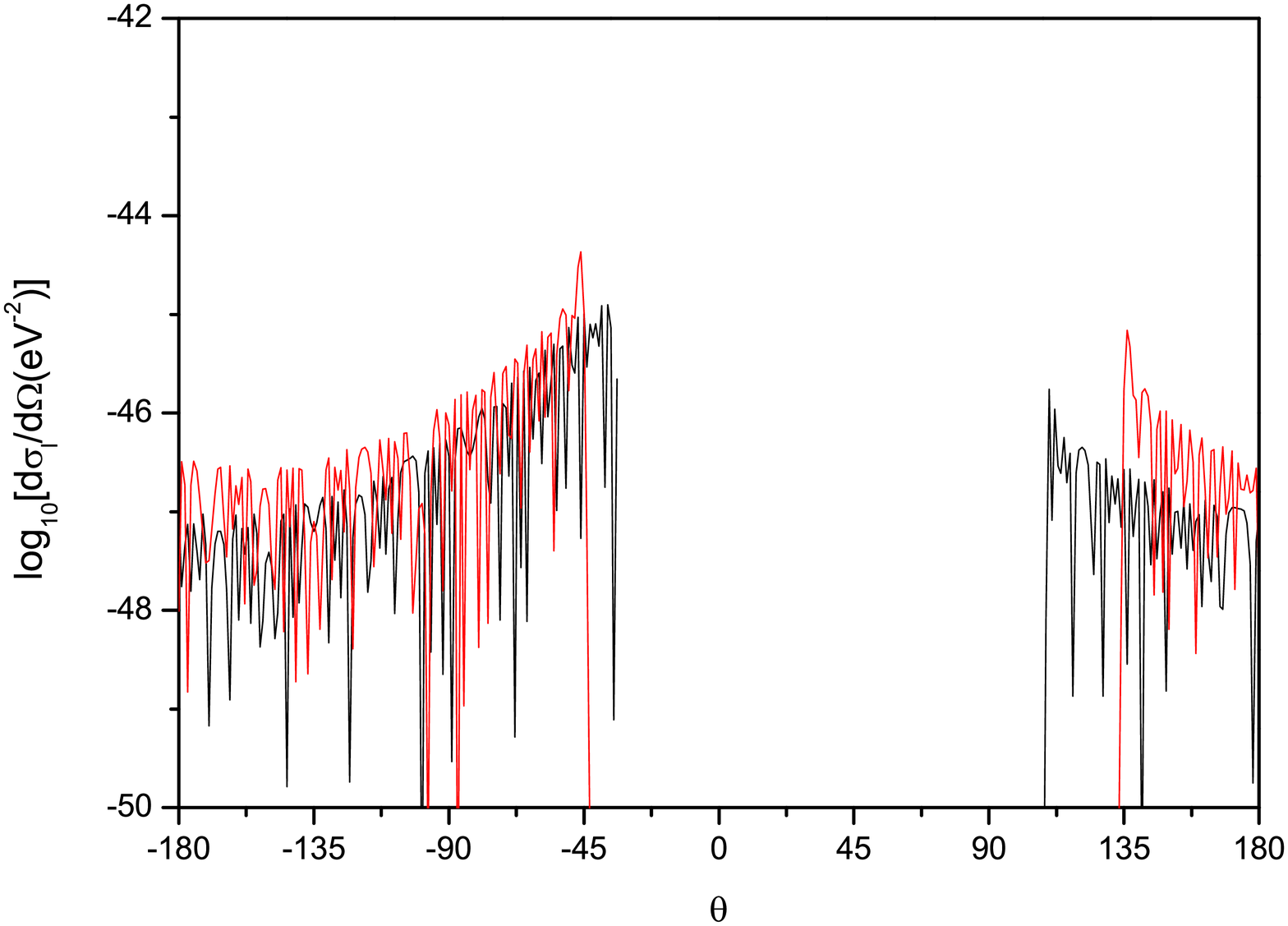}}

\caption {\label{TgL} Scattering angular distributions. The field strength has been chosen as $\varepsilon_{0}=5.18\times10^{7} V/cm$ with $\omega=1.17 ev$. The incident positron kinetic energy $T_{i}=1MeV$. The black lines represent the case for photon transfer number l = 200, while the red ones are for l = 500. The dot line stands for the laser-free case. The angle to $z$ axis of incident positron and the azimuthal angle are selected as (a)$\alpha=0^{o}, \varphi=0^{o}$; (b)$\alpha=45^{o}, \varphi=0^{o}$.}
\end{figure}
Figure 2 shows the scattering angular distributions of the partial differential cross section. In order to investigate the asymmetry phenomenon, we simulate the process choosing the polarization vector $\varepsilon_{0}$ along the $x$ axis ($\alpha=0^{o}$), the incident positron momentum parallel the wave vector $k$ along the $z$ axis, the polarization vector $\varepsilon_{0}$ perpendicular to $k$ lies in the $xy$ plane.  The symmetry relevant to the $z$ axis will hold if the partial differential cross section of spatial angular ($\theta,\varphi$) is equivalent to that of ($-\theta,\varphi$), because ($\theta,\varphi+180^{o}$) and ($-\theta,\varphi$) share the same rectangular coordinates. Fig. 2(a) shows the symmetric scattering angle distributions from $-180^{o}$ to $180^{o}$. The greater the number of photons transferred, the more visible the dark windows appear. Dark windows are visible in small angles and in big ones. These are easily comprehensible as the  generalized Bessel function drops down sharply owing to $|l| > |\xi|$. When $\alpha=0^{o}$, $\xi$ can be represented as an even function of the scattering angle as $ |(ea|q_{f}| \sin\theta)/(\omega Q_{f}-\omega|q_{f}| \cos\theta)|$ from Eq. (16), which is the reason for symmetric scattering angle distribution. When $\alpha\neq0^{o}$ the argument $\xi$ cannot be neglected, which causes the asymmetry effect in Fig. 2(b).

\begin{figure}[htp]

\centering
\subfigure[]{
\label{fig:subfig:3a}
\includegraphics[bb=37 22 732 592,width=2.6in]{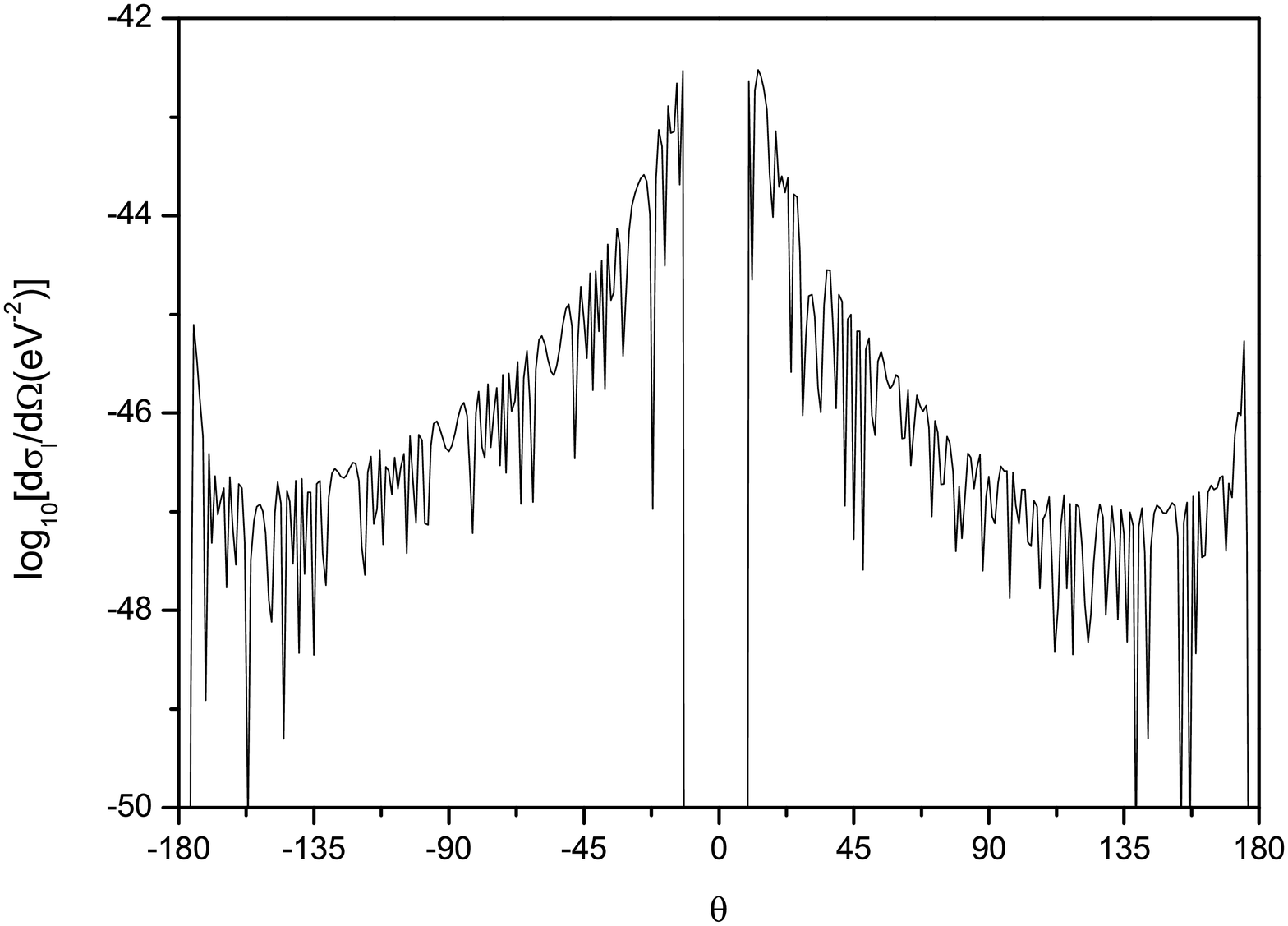}}

\hspace{0.1in}
\centering
\subfigure[]{
\label{fig:subfig:3b}
\includegraphics[bb=56 28 736 594,width=2.6in]{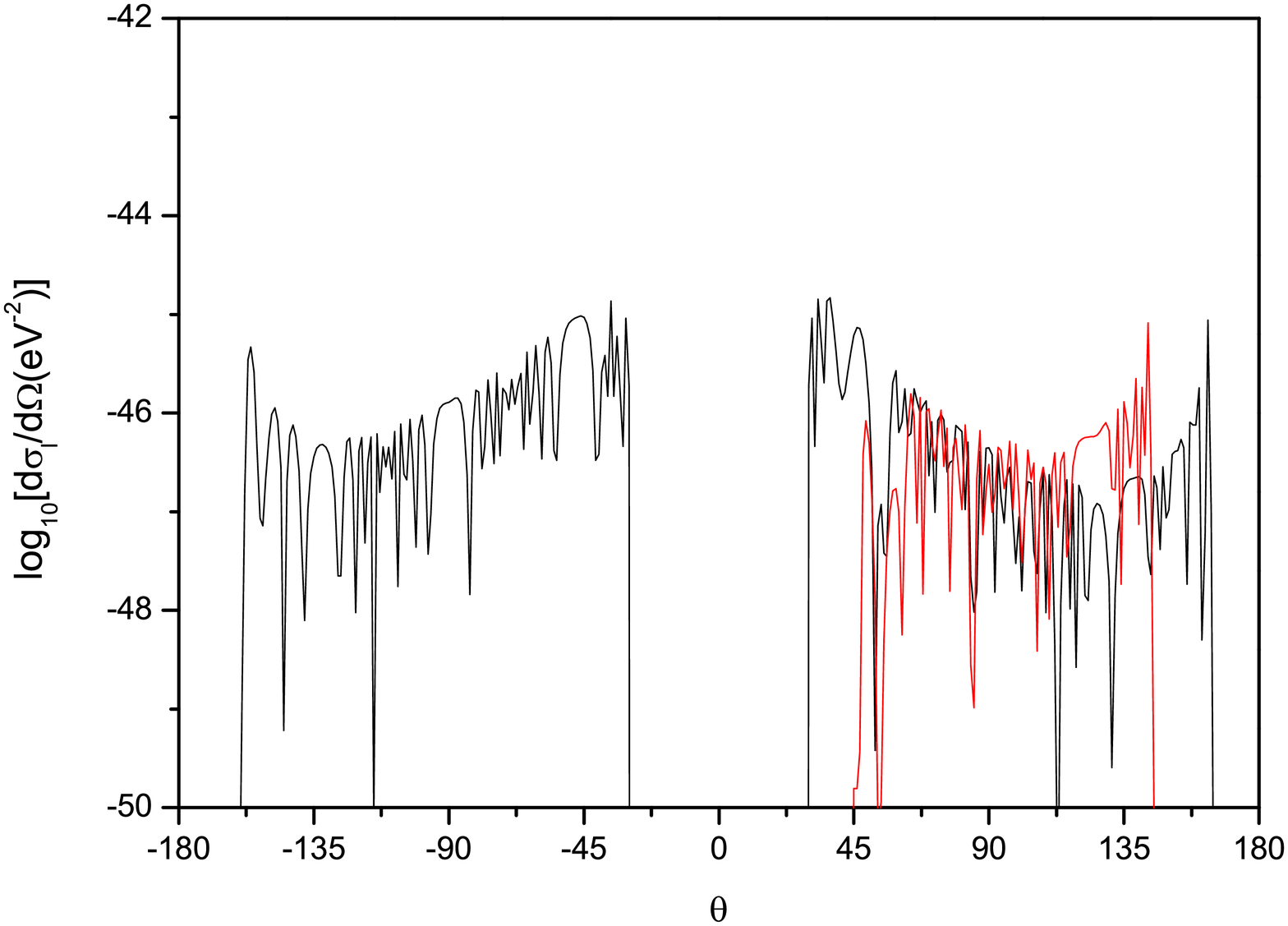}}

\caption {\label{TgL} Asymmetric scattering angular distributions. The parameters have been chosen as $\varepsilon_{0}=5.18\times10^{7} V/cm, \omega=1.17 ev, T_{i}=1MeV, \alpha=0^{o}, \varphi=30^{o}$. (a) The black lines represent the case for photon transfer number $l = 50$. (b) The black lines represent the case for photon transfer$l=200$, while the red ones are for $l = 500$.}
\end{figure}
Figure 3 shows the asymmetric scattering angular distributions versus the different photon transfer numbers. Symmetry will not hold even at $\alpha=0^{o}$ if we choose the azimuthal angle $\varphi=30^{o}$. Symmetry violation extends to a lager scale as more laser photons participate in the collision process. We can see in Fig. 3(b) that the partial differential cross section only survives on the positive side when the emission photon number is selected as $500$. The reason is that $\xi$ will not stay as an even function for its denominator changes to $\omega(Q_{f}-|q_{f}| \sin\theta \cos\varphi-|q_{f}| \sin\theta \sin\varphi- |q_{f}| \cos\theta )$.

\begin{figure}[htp]

\centering
\subfigure[]{
\label{fig:subfig:4a}
\includegraphics[bb=37 22 732 592,width=2.6in]{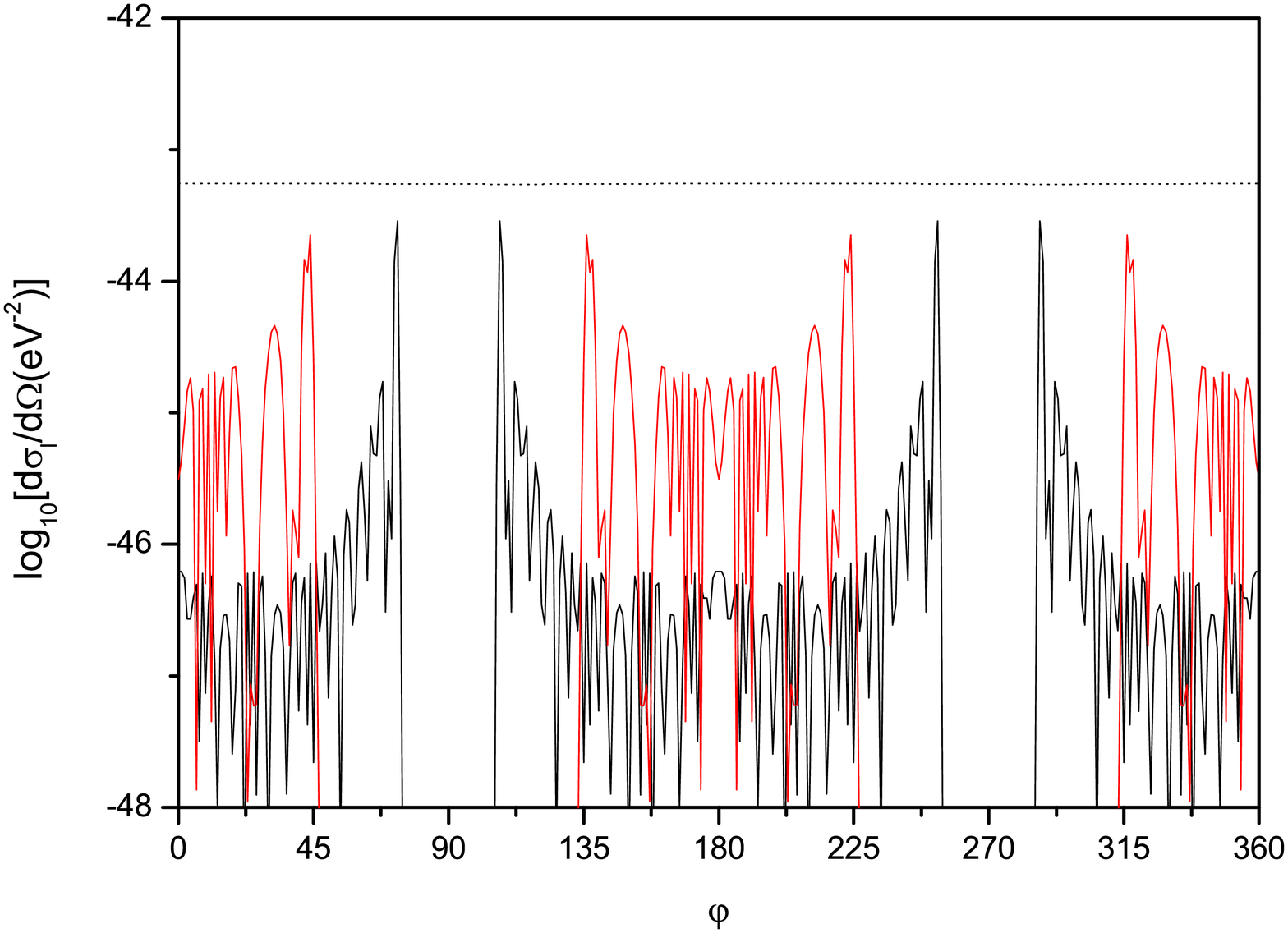}}

\hspace{0.1in}
\centering
\subfigure[]{
\label{fig:subfig:4b}
\includegraphics[bb=56 28 736 594,width=2.6in]{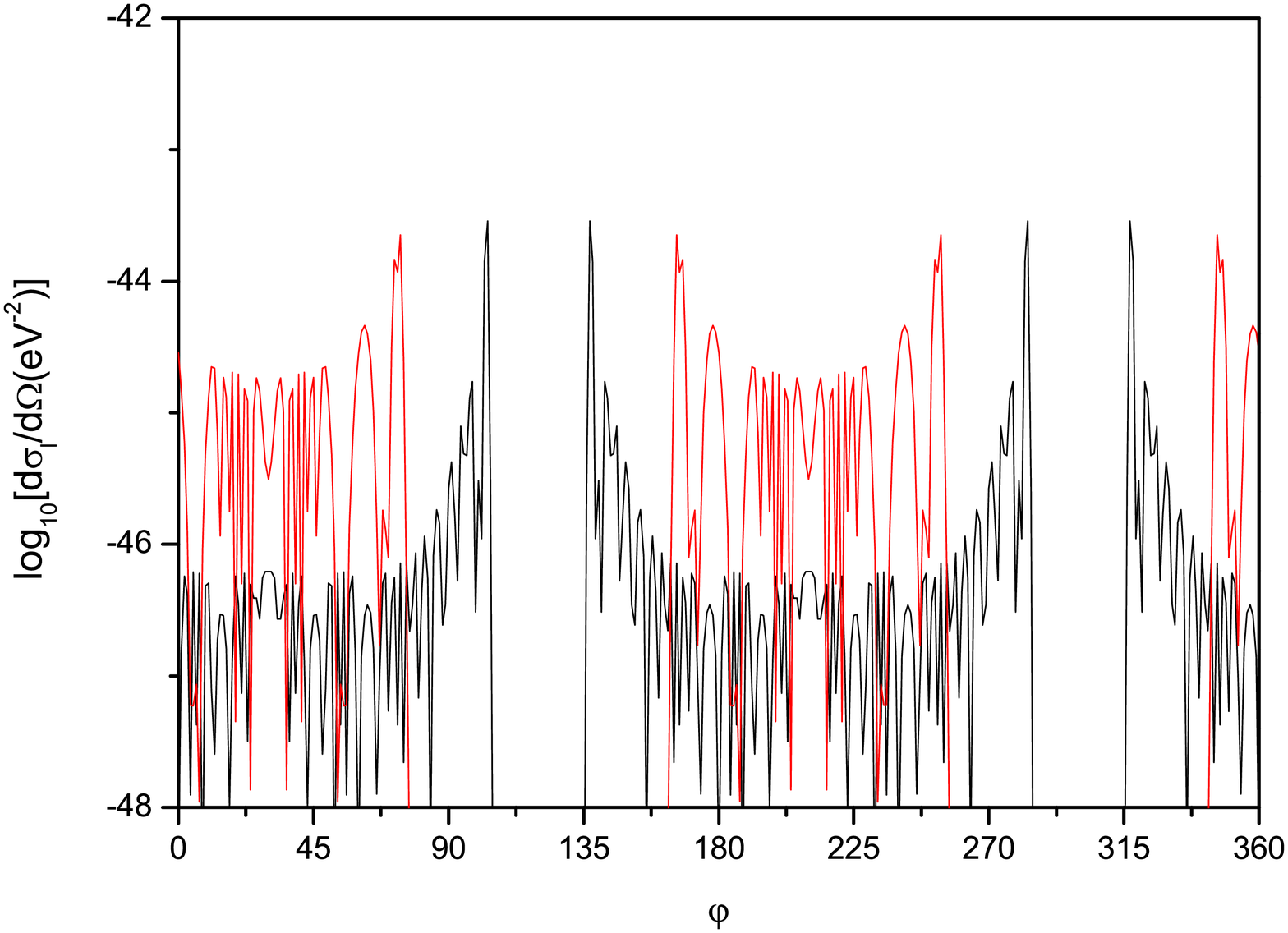}}

\hspace{0.1in}
\centering
\subfigure[]{
\label{fig:subfig:4c}
\includegraphics[bb=56 28 736 594,width=2.6in]{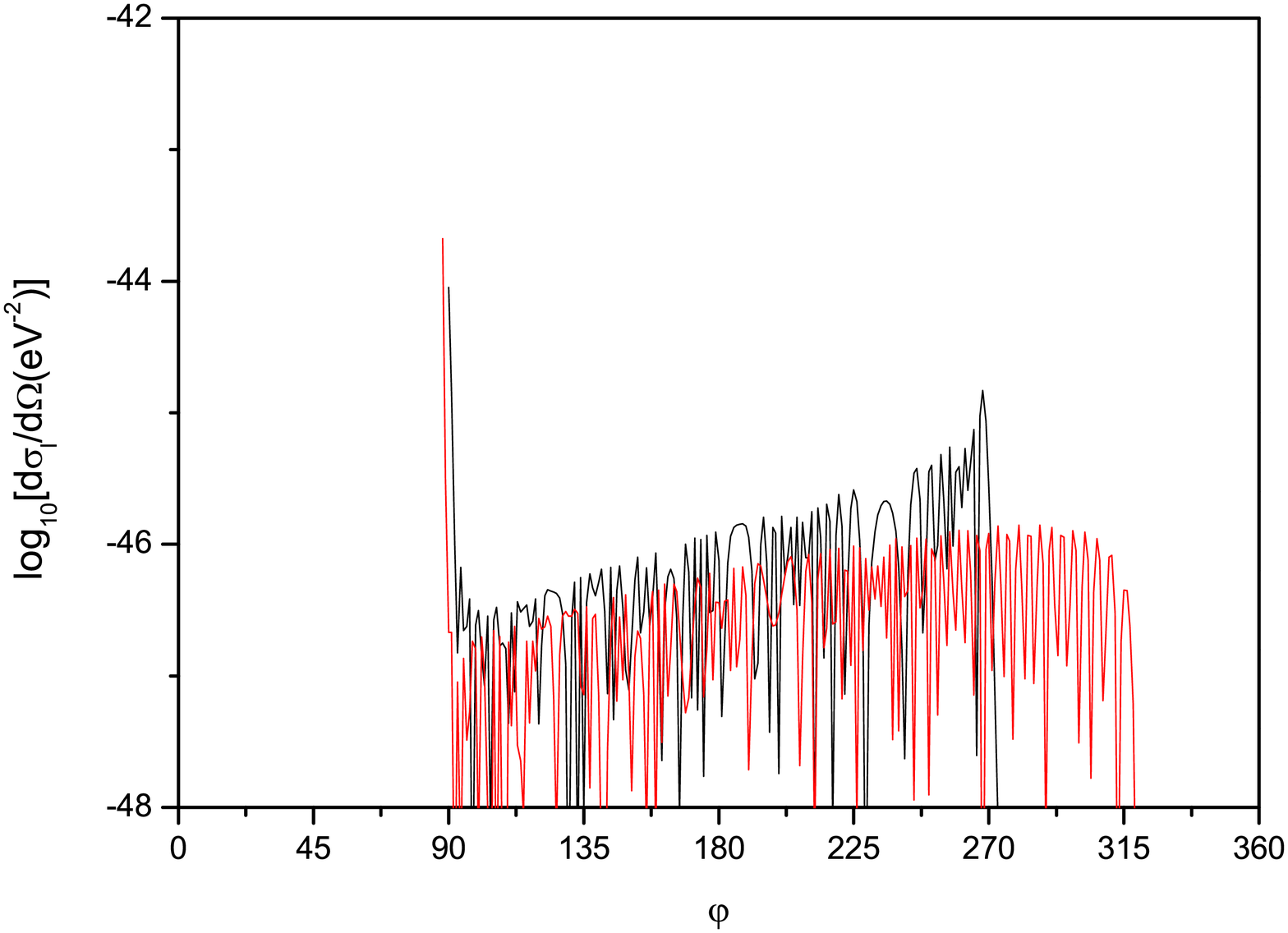}}

\caption {\label{TgL} The partial differential cross section as a function of azimuthal angle. The parameters: $\varepsilon_{0}=5.18\times10^{7} V/cm, \omega=1.17 ev, T_{i}=MeV, \theta=90^{o}$. The dot line is for the laser-free case. $k//p_{i}//z$ : (a) $\alpha=0^{o}$, black lines: $l = 200$; red lines: $l = 500$; (b) $\alpha=30^{o}$, black lines:$l = 200$; red lines: $l = 500$. $k//y$, $p_{i}//z$ : (c) $\alpha=45^{o}$, black lines: $l = 200$; red lines: $l = 500$.}
\end{figure}
Figure 4 shows the azimuthal angle distributions of the partial differential cross section. We choose the incident positron is parallel the direction of the propagation of the laser field $k$ along the $z$ axis in (a) and (b). Another symmetry violation belongs to the shift of azimuthal angle distributions originating from the rotation of the polarization vector of laser fields as shown in this picture. Intuitively the partial differential cross section can be symmetric relevant to the plane formed by the momentum vector of incident positron $p_{i}$, the laser field wave vector $k$, and the polarization vector $\epsilon$, if they keep in the same plane. The $xz$ plane is such as the symmetry plane in the case of Figs. 4(a). Symmetry still holds with some horizontal shift when the polarization vector $\epsilon$ rotates around the $z$ axis by a certain degree $\alpha$ [$30^{o}$ in Fig. 4(b)]. When we choose the laser field wave vector $k$ is along the $y$ axis and $p_{i}$ remains along the z axis, the symmetry violates in Fig. 4(c). That is because the initial three vectors fail to be coplanar, which gives rise to the symmetry-plane vanishing. To be more specific, the symmetry of the partial differential cross section depends on the symmetry of $|\xi|$, which can be written completely as
$|\xi|=|\frac{ea|q_{f}| \sin\theta \cos\varphi}{\omega (Q_{f}-|q_{f}|\sin\theta\sin\varphi-|q_{f}|\cos\theta)}+ const|$, ($const=0$ if $\alpha=0^{o}$ or $p_{i}//k$). The factors bringing about symmetry violation are ¡°$|q_{f}|\sin\theta\sin\varphi$¡± in the denominator and ¡°const¡± from the second term of $\xi$ .

\begin{figure}[htp]

\vspace*{1cm}
\centering{
\includegraphics[bb=34 37 733 586,width=2.6in]{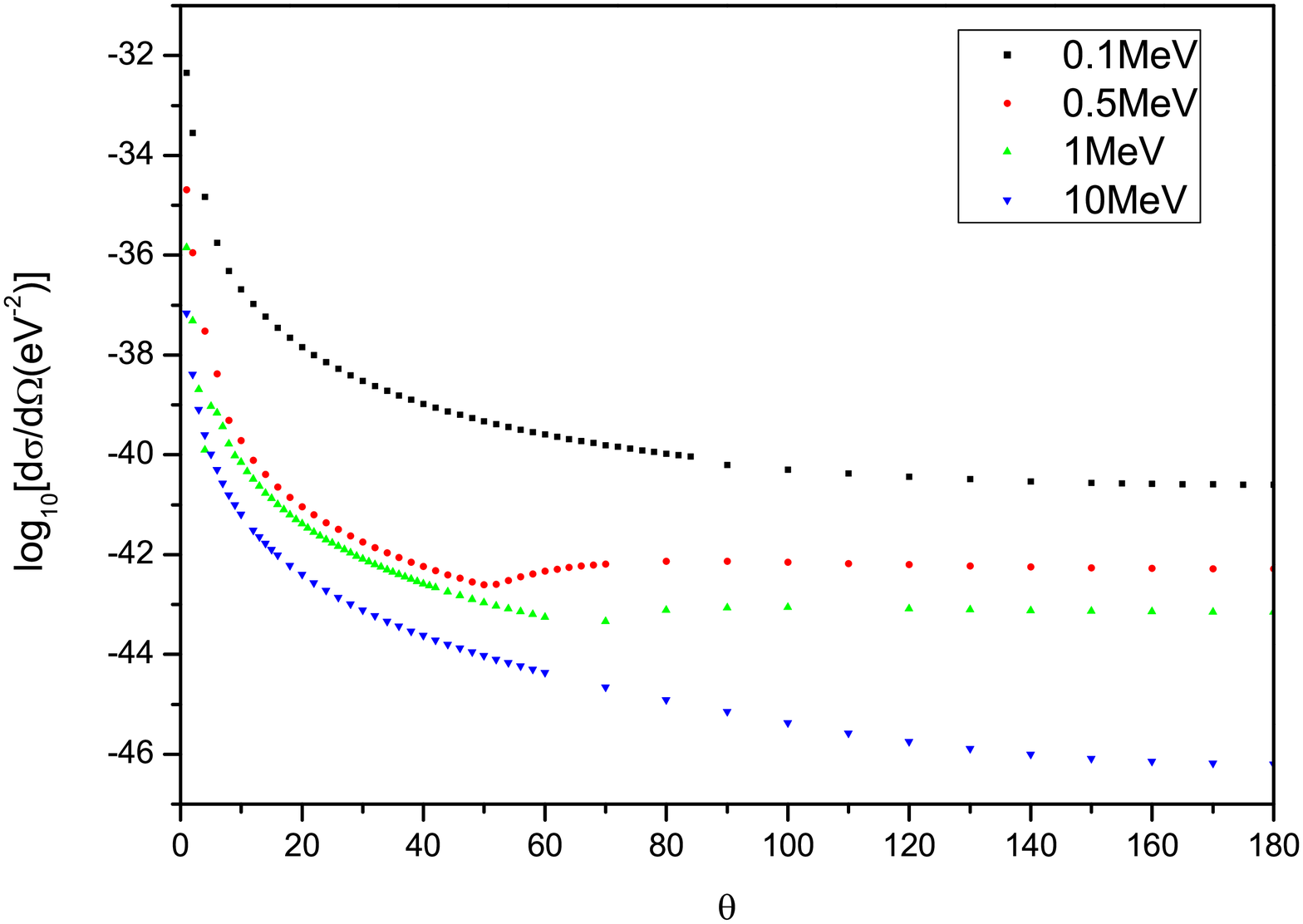}}

\caption {\label{PgL} Scattering angular distributions of the total differential cross section versus the different impact energies. The impact kinetic energies are $T_{i}=0.1, 0.5, 1, 10 MeV$. the laser parameters are the same as in Fig. 1.}
\end{figure}

Figure 5 displays the scattering angular distributions of the total cross section versus the different kinetic energies of the incident positron. Obviously, with the increase of the incident energy, the total differential cross section is reducing. In the collision process, the greater the energy of the incident particles, the smaller the impact of the target, the smaller of the total differential cross section. This phenomenon also can be explained by the Eq.(23). Fig 5 also shows the total differential cross section is reducing with the increase of the scattering angular. When the incident energy is $0.5 MeV$, there is an upward trend of the scattering cross section around $\theta=54^{o}$. That is because the dependence on the scattering angle of the formula $-2Q_i |{\bm q_f}|/Q_f+2|{\bm q_i}| \cos{\theta}$ in the Eq. (25). There is always a scattering angle that minimizes the value of this formula while the incident energy close to the mass of the positron. The same reason applies to the case of the incident kinetic energy is $1MeV$.

\section{CONCLUSIONS}
In this paper we have investigated the elastic scattering of a positron by a muon in the presence of a linearly polarized laser field and observed the phenomena of multiphoton absorption and emission. We find the photon emission processes predominant the photon absorption ones in the large scattering angle, which is sharply contrast to the case of electron-muon scattering. The reason lie in that the initial positron and muon formed a hydrogen-like intermediate transient state due to the Coulomb interaction. This transient state is not stable, positron and muon dissociated from this state, therefore emission lots of photons than absorbing. The nonlinear phenomena of cutoff, oscillation,
dark angular windows, and asymmetry are presented. We find the more number of the photons transfer with the laser field, the more obvious the dark windows are. The same phenomenon happens to symmetry violation, which results from the evanishment of the symmetry plane consisting of three coplanar vectors: the momentum vector of the incident positron $p_{i}$, the laser field wave vector $k$, and its polarization vector $\epsilon$, all of which can change the symmetry of $\xi$ as a function of scattering angle $\theta$ and azimuthal angle $\varphi$. For the total differential cross section, the result shows that the larger the incident energy is, the smaller the total differential cross section is.

\section{Acknowledgements}
This work is partially supported by National Natural Science Foundation of China (NSFC) under Grant Nos. 11275186,91024026,and FOM2014OF001, and the National Basic Research Program of China under Grants No. 2007CB925200 and No. 2010CB923301.

\end{document}